\documentclass[10pt, conference, letterpaper]{IEEEtran}
\IEEEoverridecommandlockouts
\usepackage{booktabs}
\usepackage{cite}
\usepackage{multirow}
\usepackage{svg}
\usepackage[T1]{fontenc}
\usepackage{aecompl}

\usepackage{amsmath,amssymb,amsfonts}
\usepackage{amsthm}

\newtheorem{prop}{Proposition}
\usepackage{graphicx}
\usepackage{textcomp}
\usepackage{subfigure}
\usepackage{pifont}
\usepackage[ruled,vlined]{algorithm2e}
\usepackage{threeparttable}
\usepackage{hyperref}
\usepackage{cleveref}
\Crefname{equation}{Eq.}{Eq.}
\Crefname{figure}{Fig.}{Fig.}
\Crefname{section}{Sec}{Sec}

\def\BibTeX{{\rm B\kern-.05em{\sc i\kern-.025em b}\kern-.08em
    T\kern-.1667em\lower.7ex\hbox{E}\kern-.125emX}}
    
\begin{document}

\title{Breaking Secure Aggregation: Label Leakage from Aggregated Gradients in Federated Learning}


\author{\IEEEauthorblockN{Zhibo Wang$^ {\dagger, \sharp}$, Zhiwei Chang$^ {\dagger, \sharp}$, Jiahui Hu$^ {\dagger, \sharp, \ast}$\thanks{* Jiahui Hu is the corresponding author}, Xiaoyi Pang$^ {\dagger, \sharp}$, Jiacheng Du$^ {\dagger, \sharp}$, Yongle Chen$^{\wr}$ and Kui Ren$^ {\dagger, \sharp}$}
\IEEEauthorblockA{$^\dagger $The State Key Laboratory of Blockchain and Data Security, Zhejiang University, Hangzhou, China}
\IEEEauthorblockA{$^{\sharp}$School of Cyber Science and Technology, Zhejiang University, Hangzhou, China}
\IEEEauthorblockA{$^{\wr}$College of Computer Science and Technology, Taiyuan University of Technology, China}

Email: \{zhibowang, czw1999, jiahuihu, jcdu, kuiren\}@zju.edu.cn,
 xypang@whu.edu.cn,
 chenyongle@tyut.edu.cn
}

\maketitle

\begin{abstract}
Federated Learning (FL) exhibits privacy vulnerabilities under gradient inversion attacks (GIAs), which can extract private information from individual gradients. To enhance privacy, FL incorporates Secure Aggregation (SA) to prevent the server from obtaining individual gradients, thus effectively resisting GIAs. In this paper, we propose a stealthy label inference attack to bypass SA and recover individual clients' private labels. Specifically, we conduct a theoretical analysis of label inference from the aggregated gradients that are exclusively obtained after implementing SA.  The analysis results reveal that the inputs (embeddings) and outputs (logits) of the final fully connected layer (FCL) contribute to gradient disaggregation and label restoration. To preset the embeddings and logits of FCL, we craft a fishing model by solely modifying the parameters of a single batch normalization (BN) layer in the original model. Distributing client-specific fishing models, the server can derive the individual gradients regarding the bias of FCL by resolving a linear system with expected embeddings and the aggregated gradients as coefficients. Then the labels of each client can be precisely computed based on preset logits and gradients of FCL's bias.  Extensive experiments show that our attack achieves large-scale label recovery with   100\% accuracy on various datasets and model architectures.
\end{abstract}


\section{Introduction}
Federated learning (FL) \cite{mcmahan2017communication} is a new paradigm of distributed machine learning (ML) that allows multiple clients to jointly train models while keeping their private data locally.
The server distributes the global model to clients, who train the received model locally using their private data.
The clients then send model updates (gradients or model parameters) to the server for updating the global model. 
As a privacy-preserving ML technology \cite{kairouz2021advances,zhao2018federated,bonawitz2019towards,9796594,wang2022towards,sun2021pain,sun2022profit,zhao2022fedinv}, FL provides services for various privacy-sensitive applications, including healthcare \cite{brisimi2018federated}, word prediction \cite{hard2018federated}, and finance \cite{long2020federated}.

But recent works on gradient inversion attacks (GIAs) \cite{wang2019beyond,zhu2019deep,geiping2020inverting,yin2021see,zhao2020idlg,hu2023shield,wang2023attrleaks,wainakh2021user,ma2022instance} have shown that private data can be revealed from model updates (especially gradients), thus exposing the privacy vulnerability of FL.
The primary objectives of existing gradient inversion attacks consist of data reconstruction and label restoration.
Data reconstruction attacks \cite{zhu2019deep,geiping2020inverting,yin2021see} aim to recover the original input data based on iterative gradient matching, which minimizes the distance (e.g., $L_2$ norm) between the gradients generated from original data and that generated from reconstructed data.
Specifically, most of these attacks would recover the labels of the samples in advance, which can reduce the number of iterative rounds required to recover original data. 
So the research focused on inferring the original labels has also received attention, such as iDLG\cite{zhao2020idlg}, LLG\cite{wainakh2021user} and iLRG\cite{ma2022instance}.
These attacks aim to determine whether certain labels are present in the client's training batch and the number of these labels, and they recover labels by analyzing the gradients of the final fully connected layer.

\begin{figure}[!t]
    \vspace{-2mm}

    \centering
    \includegraphics[scale=0.35]{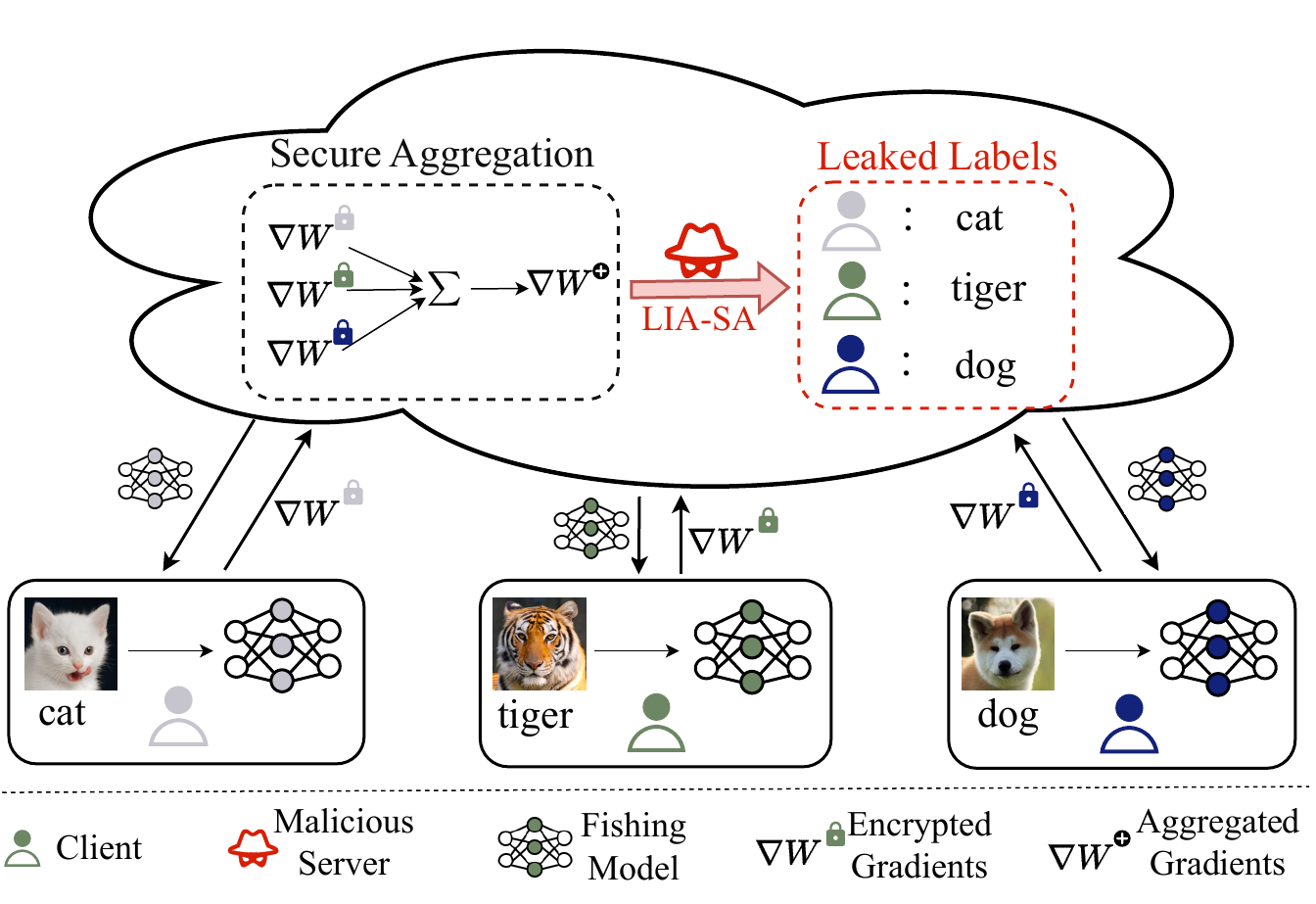}
    
    \caption{Illustration of our Label Inference Attack against SA (LIA-SA).}
    \label{fig:intro}
    \vspace{-6mm}
\end{figure}

However, the above gradient inversion attacks require access to the gradients provided by individual client and would be prevented if SA \cite{bonawitz2017practical} is applied to FL. SA is a multi-party computation (MPC) protocol that enables secure computation of the sum of several private inputs, thereby preventing the server from accessing individual gradients and enhancing the security of FL. 
Recent GIAs \cite{pasquini2022eluding, zhao2023secure} have studied the privacy leakage problem in FL with SA, revealing the vulnerability of SA.
The authors in \cite{pasquini2022eluding} distribute a normal model to a target client and modified models to other non-target clients. Those modified models will produce zero gradients. After secure aggregation, the individual gradients of the target client are exposed to the server.
\cite{zhao2023secure} reconstructs private data from all clients by inserting a module of a few layers at the start of the original model. 
Both works only aim at data reconstruction while ignoring label restoration and lacking stealthiness because abnormal modifications to the model architecture \cite{zhao2023secure} or all-zero gradients \cite{pasquini2022eluding} are easily detected by clients.

To this end, we propose a \textbf{L}abel \textbf{I}nference \textbf{A}ttack against \textbf{S}ecure \textbf{A}ggregation, called LIA-SA, which can break the protection of SA and recover the labels of clients from aggregated gradients. 
Moreover, our attack performs well on stealthiness because it does not modify the model architecture and only modifies a single layer's parameters. Specifically, we first conduct a theoretical analysis on label inference from aggregated gradients, which discloses that the inputs (embeddings) and outputs (logits) of the final fully connected layer (FCL) \footnote{The FCL mentioned in this paper refers to the final fully connected layer.} contribute to gradient disaggregation and label restoration. That is, we can derive the gradient information of individual clients and recover their private labels if we have the embeddings and logits of each client. However, the embeddings and logits are unable to be captured in the benign FL process and vary with the input data.  To make the embeddings and logits data-agnostic and predetermined, we assume a malicious server and design a fishing model by modifying the parameters of one batch normalization (BN) layer in the global model. During the process of distributing the global model, the malicious server replaces the global model with client-specific fishing models, i.e., different fishing models distributed to different clients. After receiving the aggregated gradients, the malicious server can derive the individual gradients of FCL's bias by resolving a linear system with expected embeddings and the aggregated gradients as coefficients. Then the labels of each client can be precisely computed based on preset logits and individual gradients of FCL's bias.
\Cref{fig:intro} illustrates our proposed LIA-SA algorithm.
 
Our main contributions are as follows:

\begin{itemize}
\item We propose a novel label inference attack LIA-SA that infers the private labels of each client from the aggregated gradients. This further demonstrates that SA offers a false sense of privacy protection in FL.

\item  We theoretically analyze the label leakage from aggregated gradients and design fishing models to achieve the perfect label inference attack against SA. Compared to existing gradient inversion attacks against SA, the LIA-SA conducts fewer modifications to model parameters and presents better stealthiness.
\item Extensive experiments show that LIA-SA achieves large-scale label recovery with  100\% accuracy on various datasets and model architectures.  Moreover, LIA-SA  also restores a certain amount of labels when additional defenses are implemented.
\end{itemize}


\section{Related Work}\label{sec:related_work}
In this section, we introduce some works related to GIAs,
which are divided into passive attacks and active attacks according to the attacker's capabilities.
\subsection{Passive Attacks}
Passive attacks\cite{boenisch2021curious} assume that the attacker is honest but curious, which means that the attacker can only launch an attack by observing the received gradients and cannot maliciously manipulate the FL protocol (i.e., modify the model architecture or parameters).

\textbf{Data Reconstruction Attacks} aim to recover the private inputs from gradients.
A study by Wang et al.\cite{wang2019beyond} inverses single image representation from gradients generated by a 4-layer CNN.
They use GAN to generate discrimination on client identity and realize an attack on user-level privacy.
Zhu et al.\cite{zhu2019deep} propose Deep Leakage from Gradients (DLG), which optimizes dummy data and labels by gradient matching and achieves pixel-level recovery of images.
Geiping et al.\cite{geiping2020inverting} improve the loss function in DLG from Euclidean distance to cosine similarity, which realizes image restoration at high resolution.
While the above attacks work well on a single image, they don't perform well on batches.
GradInversion\cite{yin2021see} achieves success in the reconstruction of mini-batches with the help of BN statistics.

\textbf{Label Inference Attacks} are intended to infer labels from gradients.
Unlike the data reconstruction attacks, most label inference attacks only utilize the gradients of FCL.
Zhao et al.\cite{zhao2020idlg} present an analytical approach, called iDLG, to extract the ground-truth label from the shared gradients.
By observing the sign of the gradients of FCL, they successfully infer the ground-truth label of a single image.
In order to deal with multiple samples, Yin et al.\cite{yin2021see} propose a batch label restoration method, which extends the label inference attack from a single sample to a mini-batch.
However, these attacks are only applicable when there are no two samples in the batch belonging to the same class.
Subsequent works break this restriction and enable the label inference attack to a large batch.
Based on observations of the direction and magnitude of the gradients, Wainakh et al.\cite{wainakh2021user} propose a label inference attack that can determine the number of labels in a batch.
Ma et al. propose Instance-wise Batch Label Restoration via Gradients (iLRG) \cite{ma2022instance}, which is the state-of-the-art label inference attack.
They approximately recover the average post-softmax probability and establish a system of linear equations to derive the number of labels of each class.

Passive GIAs have made remarkable progress, while they are only applicable when the server has access to gradients of the victim clients.
Once SA is applied in the FL system, the above attacks will be completely invalidated.
Although some passive attacks \cite{lam2021gradient,kariyappa2023cocktail,marchand2023sratta} can recover data from aggregated gradients, they make impractical assumptions (e.g., additional user participation information in \cite{lam2021gradient}, fully connected networks in \cite{kariyappa2023cocktail,marchand2023sratta}) and have very narrow applications.

\subsection{Active Attacks}
Passive attacks assume that the server is honest but curious, which greatly underestimates the server's ability to launch attacks. Recently, researchers have begun to focus on active attacks\cite{boenisch2021curious,pasquini2022eluding,fowl2021robbing,wen2022fishing,zhao2023secure,zhao2023resource}, where
a malicious server can make full use of its capabilities to launch an active attack in order to recover as much private data as possible.

Initially, active attacks focus on individual user gradients, i.e., the server has access to the individual gradients of each user.
Fowl et al.\cite{fowl2021robbing} add a large fully connected layer at the front of the original model to separate the gradients of different samples.
This model modification strategy can extract data from arbitrarily large batches as long as there are enough neurons in the fully connected layer.
The attack proposed in \cite{wen2022fishing} only actively modifies the model parameters and successfully recovers a single sample from the large batches.
Boenisch et.al\cite{boenisch2021curious} create trap weights by initializing the weights of the fully connected layer as half positive and half negative.
The trap weights allow the gradients of different samples to be separated into different neurons, and the image can be perfectly reconstructed through the gradients of the fully connected layer.

At the same time, some works are also studying active attacks against SA.
In \cite{zhao2023secure,zhao2023resource}, the authors add a convolutional layer and two fully connected layers in the front of the original model to separate gradients from different clients and different samples. They recover a large number of samples from aggregated gradients.
Pasquini et al.\cite{pasquini2022eluding} send the normal model to a target client while sending malicious models to the other non-target clients.
The malicious models force non-target clients to produce almost zero gradients, which makes SA no longer work and exposes the gradients of the target client.
However, existing GIAs against SA only focus on data reconstruction and lack stealthiness.
For example, the model modification in \cite{zhao2023secure,zhao2023resource} and the all-zero gradients in \cite{pasquini2022eluding} are easily detected by the clients.
Therefore, the goal of our attack is to break the protection of SA, recover the labels from the aggregated gradients, and ensure stealthiness.

\section{Preliminary}\label{sec:preliminary}
In this section, we provide some relevant background knowledge, including federated learning, secure aggregation, and our threat model.

\subsection{Federated Learning}
Federated learning (FL) is a new distributed machine learning paradigm that allows a set of clients $\mathcal{N}$ to jointly train a global model without sharing their private data. The training process of FL requires multiple rounds until the global model converges or achieves a given accuracy. For each round, $U$ randomly selected clients  $\mathcal{U}=\{1,\dots,U\}$ $(\mathcal{U} \subseteq \mathcal{N})$ participate in FL. 
Each client $u \in \mathcal{U}$ holds a private training dataset $\mathcal{D}_u = \left\{ (\boldsymbol{x}^u(1),c^u(1),\dots,(\boldsymbol{x}^u(D_u),c^u(D_u)) \right\}$, including $D_u$ data samples (i.e., $|\mathcal{D}_u|  = D_u$). $\boldsymbol{x}^u(k)$ and $c^u(k)$ denote the $k^{th}$ sample and the corresponding label in client $u$.

There are various FL algorithms, such as federated stochastic gradient descent (FedSGD) and federated averaging (FedAVG)\cite{mcmahan2017communication}.  In this paper, we mainly focus on  FedSGD, which is a typical FL algorithm commonly assumed in GIA studies\cite{zhu2019deep,zhao2020idlg,wen2022fishing,wainakh2021user,ma2022instance}. At round $t$, the server sends global model parameters $\boldsymbol{W}^t$ to clients $\mathcal{U}$. Then, each client $u \in \mathcal{U}$ uses this global model to perform local training based on local dataset $\mathcal{D}_u$ and sends back its gradients $\nabla \boldsymbol{W}_u^t$ to the server for aggregation. In particular, the gradient $\nabla \boldsymbol{W}_u^t$ is derived from one-step stochastic gradient descent (SGD) based on a batch of samples in $\mathcal{D}_u$, represented as $\mathcal{B}_u$. That is, 

\begin{equation}\label{eq:gradient} 
    \nabla \boldsymbol{W}_u^t = \frac{1}{|\mathcal{B}_u|} \sum_{(\boldsymbol{x},c) \in \mathcal{B}_u} \nabla \mathcal{L} (\boldsymbol{x},c;\boldsymbol{W}^t),
\end{equation}
where $\mathcal{L}(\boldsymbol{x},c;\boldsymbol{W}^t)$ represents the loss function associated with the data sample $(\boldsymbol{x},c)$ and the model parameters $\boldsymbol{W}^t$.
After receiving the gradients from all selected clients, the server updates the global model as follows:
\begin{equation}\label{eq:aggregation} 
    \boldsymbol{W}^{t+1}=\boldsymbol{W}^{t}-\eta \frac{\sum_{u \in \mathcal{U}} \nabla \boldsymbol{W}_u^t} {\left\lvert \mathcal{U} \right\rvert},
\end{equation}
where $\eta$ is the global learning rate and $\left\lvert \mathcal{U} \right\rvert = U$ is the total number of selected clients at round $t$.

\subsection{Secure Aggregation}\label{sec:sa}

Secure Aggregation (SA) \cite{bonawitz2017practical} is a multi-party computation protocol designed to securely compute the sum of inputs from multiple clients without revealing the individual inputs of each client. In particular, in the FL system with a SA protocol \cite{kadhe2020fastsecagg,bell2020secure,so2022lightsecagg}, each client $u$ encrypts its own gradients $\mathrm{E}(\nabla \boldsymbol{W}_u)$ (i.e., SA encoder) before sending them to the server. After receiving the encrypted gradients, the server can decode the sum of gradients as follows:
\begin{equation}\label{eq:sa} 
    \mathrm{D}(\sum_{u \in \mathcal{U}} \mathrm{E}(\nabla \boldsymbol{W}_u)) = \sum_{u \in \mathcal{U}} {\nabla \boldsymbol{W}_u} = \nabla \boldsymbol{W}_{sa}, 
\end{equation}
where $\nabla \boldsymbol{W}_{sa}$ denotes the aggregated gradients and $\mathrm{D}(\cdot)$ is the SA decoder.
Intuitively, SA significantly enhances the privacy of FL by preventing other parties (e.g., the server) from accessing individual gradients and disrupting the linkage between gradients and clients (i.e., providing anonymity). Moreover, since the server only obtains aggregated gradients, conducting GIAs  to infer the client's private data by analyzing individual gradients becomes challenging.

\begin{figure*}[!t]
    \centering
    \includegraphics[scale=0.32]{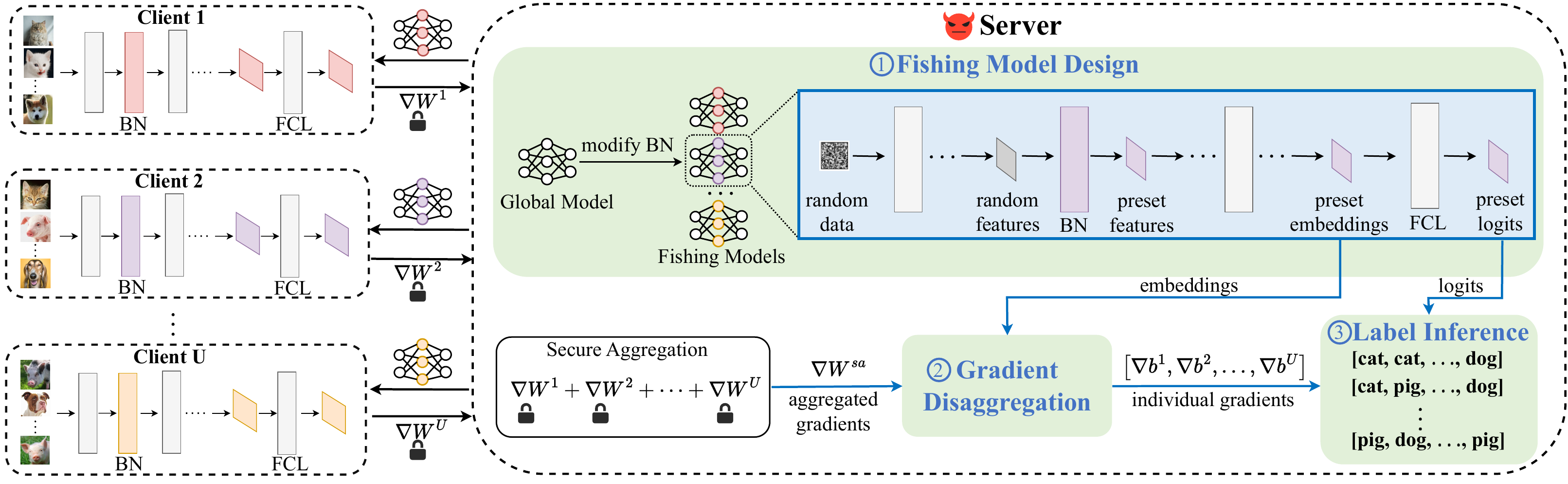}
    \caption{Overview of the proposed Label Inference Attack against SA (LIA-SA).}
    \label{fig:overview}
\end{figure*}

\subsection{Threat Model}
We consider an FL system that collaboratively trains a classification model that contains at least one fully connected layer as the last layer and uses a softmax activation with cross-entropy loss, such as ResNet\cite{he2016deep} and VGG\cite{simonyan2014very}.
The FL system employs the SA protocol to enhance the protection of data privacy.

In this paper, we assume a malicious central server $S$ and multiple honest clients $\mathcal{U}$. Following prior works \cite{pasquini2022eluding,wen2022fishing,boenisch2021curious}, the malicious server can modify the global model and distribute it to clients for local training.
To ensure stealthiness (i.e., making it difficult for clients to detect abnormalities), the server employs the following actions. Firstly, the server will maintain the original architecture of the global model, as any modification to the architecture is easily noticeable (e.g. adding a fully connected layer in front of the model\cite{fowl2021robbing,zhao2023secure}). Secondly, the server makes trivial modifications to the model parameters, ensuring that the modified model parameters and corresponding gradients remain similar to those observed under benign training conditions. As for the clients, they honestly perform the local training and compute the gradients on a batch of private data following the FedSGD. Then the gradients are encrypted using the SA encoder and submitted to the server for aggregation.
Based on the SA decoder, the malicious server can get the aggregated gradients $\nabla \boldsymbol{W}_{sa}$ but cannot access the individual gradients.

Sincerely, the objective of the malicious server is to break the protection of SA and steal the labels of clients' training data, i.e., obtaining the occurrences of each label in batch $\mathcal{B}$ from each client. Moreover, the server aims to associate the leaked labels with specific clients, which further exacerbates the privacy issue by destroying anonymity.

\section{Approach}\label{sec:method}
In this section, we introduce the proposed Label Inference Attack against SA (LIA-SA).
In Sec IV-A, we provide an overview of the LIA-SA.
In Sec IV-B, we give a theoretical analysis of recovering labels from aggregated gradients and then introduce the design of fishing models to achieve label inference in Sec IV-C.

\subsection{Overview of LIA-SA}
The goal of LIA-SA is to bypass SA and infer private labels from the aggregated gradients. By theoretically analyzing the label recovery from the aggregated gradients (illustrated in Sec IV-B), we discover that the inputs (embeddings) and outputs (logits) of FCL contribute to gradient disaggregation and label inference. However, the inputs and outputs of any intermediate layer in the client model are kept locally and difficult for the server to access. Moreover, those intermediate values are dependent on the samples, thus hard to estimate by external parties (i.e., the malicious server). Both factors present non-trivial challenges for the server to disaggregate the received gradients and infer private labels in the clients' training data.

Motivated by these challenges, we design the fishing model (illustrated in Sec IV-C), which enables the server to obtain the embeddings and logits of FCL. The intuition of designing fishing models is to make the embeddings and logits of FCL data-agnostic and only determined by model parameters by modifying the parameters of a single layer (e.g. BN layer) in the original model. In this context, the server can send fishing models with diverse parameters to different clients, producing diverse embeddings and logits for different clients. Note that all of these values are known to the server as it can input random data to fishing models to obtain the desired preset embeddings and logits.

As shown in \Cref{fig:overview}, the malicious server initiates the proposed LIA-SA by designing $U$ (i.e., the number of selected clients participating in the current round) fishing models. Then, the server generates the corresponding embeddings and logits of FCL by performing forward propagation on the fishing models using random data. Following the FL process, the server distributes those fishing models to $U$ clients for local training. After receiving the encrypted gradients and computing the aggregated gradients with the SA decoder, the server proceeds to disaggregate the aggregated gradients to obtain the individual gradients of FCL's bias (i.e., $\nabla \boldsymbol{b}$) based on the preset embeddings. Finally, the server can successfully recover the labels of each client based on the preset logits and derived individual gradients of FCL's bias.

\subsection{Label Inference from Aggregated Gradients}
In this section, we theoretically analyze how to infer labels from the aggregated gradients encrypted by SA. Inspired by \cite{zhao2020idlg} and \cite{wainakh2021user}, we begin by introducing the label inference from one-sample gradients and then proceed to discuss the label inference from batch-averaged gradients. These preliminary analyses will support and help to understand the label recovery from aggregated gradients.

\textbf{Label Inference from One-Sample Gradients.}
Suppose $\boldsymbol{\mathcal{W}}$ $\in$ $\mathbb{R}^{n \times m}$ and $\boldsymbol{b} \in \mathbb{R}^{n \times 1}$ refer to the weight and bias of FCL, and $\boldsymbol{e} \in \mathbb{R}^{m \times 1}$ denote the input embedding vector, the forward pass on such FCL can be written as $\boldsymbol{y} = \boldsymbol{\mathcal{W}} \boldsymbol{e} + \boldsymbol{b}$. $\boldsymbol{y} = \left[y_1,\dots,y_n\right]$ is the output vector of the FCL, and $n$ represents the number of total classes.
Hence, the loss value of sample $(\boldsymbol{x},c)$ on the model $\boldsymbol{W}$ is
\begin{equation}\label{eqn5} 
    \mathcal{L} (\boldsymbol{x},c;\boldsymbol{W}) = - \sum_{i=1}^{n} \hat{y}_i \ln{(\boldsymbol{softmax} (y_i))} = -\ln{\frac{e^{y_c}}{\sum_{j=1}^{n}e^{y_j}}},
\end{equation}
where $[\hat{y}_1,\hat{y}_2,\dots,\hat{y}_n]$ is the one-hot vector of the ground-truth label and $\hat{y}_c = 1$ (i.e., $\hat{y}_i = 0$, if $i \neq c$).
We note that the gradient of the loss $\mathcal{L}$ w.r.t. the output $y_i$ is
\begin{eqnarray}
    \frac{\partial{\mathcal{L}}(\boldsymbol{x},c;\boldsymbol{W})}{\partial{y_i}} &=&-(\frac{\partial{\ln{e^{y_c}}}}{\partial{y_i}} -
            \frac{\partial{\ln{\sum_{j=1}^{n} e^{y_j}}}}{\partial{y_i}}) \nonumber \\
    ~&=&\begin{cases}
            -1+ \frac{e^{y_i}}{\sum_{j=1}^n e^{y_j}}    & i=c\\
            \frac{e^{y_i}}{\sum_{j=1}^n e^{y_j}}    & i\neq c.\\
        \end{cases}
    \label{eqn6} 
\end{eqnarray}
Let $\boldsymbol{\mathcal{W}}_i$ and $b_i$ denote the weight and bias of the $i^{th}$ neuron in FCL, respectively.
According to the chain role, we have $\frac{\partial{y_i}}{\partial{b_i}} = \frac{(\partial{\boldsymbol{\mathcal{W}}_i\boldsymbol{e}+b_i})}{\partial{b_i}} = 1$.
So the gradient of $\mathcal{L}$ w.r.t $b_i$ is
\begin{equation}\label{eqn7} 
    \nabla{b}_i = \frac{\partial{\mathcal{L}}(\boldsymbol{x},c;\boldsymbol{W})}{\partial{b}_i} = \frac{\partial{\mathcal{L}}(\boldsymbol{x},c;\boldsymbol{W})}{\partial{y}_i} \cdot \frac{\partial{y}_i}{\partial{b}_i} = \frac{\partial{\mathcal{L}}(\boldsymbol{x},c;\boldsymbol{W})}{\partial{y}_i}.
\end{equation}
As $\frac{e^{y_i}}{\sum_{j=1}^n e^{y_j}} \in (0,1)$, we have $\nabla{b}_i \in (-1,0)$ if $i=c$,  and  $\nabla{b}_i \in (0,1)$ if $i \neq c$.
Therefore, we can infer the ground-truth label of one sample by observing the index corresponding to the negative value in the gradients of FCL's bias $\nabla{\boldsymbol{b}}$. We summarize the above analysis as Proposition 1.
\begin{prop}
 The label $i$ is presented in the training data when the one-sample gradients of FCL's bias $\nabla{b}_i \leq 0$, where $\nabla{b}_i \in \nabla{\boldsymbol{b}}$.
\end{prop}

\textbf{Label Inference from Batch-Averaged Gradients.}
The above analysis is only applicable to a single sample. When the gradients are generated by a batch of samples (a more common scenario), a negative value may not appear in $\nabla{\boldsymbol{b}}$. Naturally, it is impossible to directly infer the label information based on the sign of $\nabla{\boldsymbol{b}}$.

In the following, we carefully analyze label inference from the batch-averaged gradients.
Given a batch of samples $\mathcal{B}=\left\{ (\boldsymbol{x}(1),c(1)),\dots,(\boldsymbol{x}(B),c(B)) \right\}$ ($B$ denotes batch size),
we use $\boldsymbol{x}(k)$ to denote the $k^{th}$ sample in the batch $\mathcal{B}$ and $c(k)$ is the ground-truth label of $\boldsymbol{x}(k)$.
Moreover, $\boldsymbol{e}(k)$ and $\boldsymbol{y}(k)$ are the input and output of FCL on the $k^{th}$ sample.
Since there are multiple samples in the batch, the gradient of FCL's bias $\nabla{b}_i \in \nabla{\boldsymbol{b}}$ can be produced as follows:
\begin{equation}\label{eqn8} 
    \nabla{b}_i = \frac{1}{B} \sum_{k=1}^{B} \frac{\partial{\mathcal{L}(k)}}{\partial{b}_i} = \frac{1}{B} \sum_{k=1}^{B} \frac{\partial{\mathcal{L}(k)}}{\partial{y_i(k)}}.
\end{equation}
For simplicity, we use $\mathcal{L}(k)$ to denote $\mathcal{L}(\boldsymbol{x}(k),c(k);\boldsymbol{W})$.
Combining Eq. \eqref{eqn6} and \Cref{eqn8}, we have
\begin{eqnarray}\label[equation]{eqn9} 
    \nabla{b}_i &=&-\frac{1}{B}\sum_{k=1}^{B}(\frac{\partial{\ln{e^{y_c(k)}}}}{\partial{y_i(k)}} -
            \frac{\partial{\ln{\sum_{j=1}^{n} e^{y_j(k)}}}} {\partial{y_i(k)}}) \nonumber \\
        &=& -\frac{\lambda_i}{B} + \frac{1}{B} \sum_{k=1}^{B} \frac{e^{y_i(k)}} {\sum_{j=1}^{n} e^{y_j(k)}},
\end{eqnarray}
where $\lambda_i$ is the number of occurrences of  label $i$ in the batch $\mathcal{B}$, and
$y_i(k)$ corresponds to the output of the $k^{th}$ sample on the $i^{th}$ neuron of FCL. $y_c(k)$ represents the output of the $k^{th}$ sample on a specific neuron of FCL,  of which the index refers to the ground-true label of the $k^{th}$ sample.

According to Eq. (\ref{eqn9}), we know that if the server has access to the gradients of bias $\nabla{\boldsymbol{b}}^u$ and the output (i.e., logits) $\boldsymbol{y}^u$ of FCL on client $u$, then it can accurately recover the labels in client's batch $\mathcal{B}^u$. We illustrate this in Proposition 2.
\begin{prop}
  The number of occurrences $\lambda_i$ of label $i$ in the training batch $\mathcal{B}$  can be derived from the batch-averaged gradients of bias $\nabla{b}_i$ and logits $\boldsymbol{y}$ of FCL.
\end{prop}

Although Proposition 2 provides guidance for inferring label information from the batch-averaged gradients, the inaccessible logits $\boldsymbol{y}$ makes label recovery from batch-averaged gradients challenging. Existing label inference attacks \cite{wainakh2021user, ma2022instance} mainly focus on this challenge and have successfully recovered labels from such batch-averaged gradients if accessing individual gradients. 
However, when FL is incorporated with SA, the server can only access aggregated gradients, not the gradients of individual client. How to infer labels from aggregated gradients remains an open problem.

\textbf{Label Inference from Aggregated Gradients.}
Here, we theoretically analyze the label inference from aggregated gradients, which includes two steps: gradient disaggregation and label inference.

\emph{Step 1: Gradient Disaggregation.}
Let $\boldsymbol{\mathcal{W}}_i$ represent the $i^{th}$ row in the weight $\boldsymbol{\mathcal{W}}$ of FCL.
The gradients of $\boldsymbol{\mathcal{W}}_i$ is
\begin{equation}\label{eqn10} 
    \nabla \boldsymbol{\mathcal{W}}_i = \frac{1}{B} \sum_{k=1}^{B} \frac{\partial{\mathcal{L}(k)}}{\partial{\boldsymbol{\mathcal{W}}_i}} = \frac{1}{B} \sum_{k=1}^{B} \frac{\partial{\mathcal{L}(k)}}{\partial{y_i(k)}} \cdot \frac{\partial{y_i(k)}}{\partial{\boldsymbol{\mathcal{W}}_i}}
\end{equation}
For the FCL, we have
\begin{equation}\label{eqn11}
    \frac{\partial{y_i(k)}}{\partial{\boldsymbol{\mathcal{W}}_i}} = \frac{\partial{(\boldsymbol{\mathcal{W}}_i \boldsymbol{e}(k)+b_i)}}{\partial{\boldsymbol{\mathcal{W}}_i}} = \boldsymbol{e}(k)^\top
\end{equation}
So combining \Cref{eqn10} and \Cref{eqn11}, we have
\begin{equation}\label{eqn12}
    \nabla \boldsymbol{\mathcal{W}}_i = \frac{1}{B} \sum_{k=1}^{B} \frac{\partial{\mathcal{L}(k)}}{\partial{\boldsymbol{\mathcal{W}}_i}} = \frac{1}{B} \sum_{k=1}^{B} \frac{\partial{\mathcal{L}(k)}}{\partial{y_i(k)}} \cdot \boldsymbol{e}(k)^\top
\end{equation}
Suppose that each sample in the batch $\mathcal{B}$ makes equal inputs to FCL, i.e., $\boldsymbol{e}(k) = \boldsymbol{e}$ for $ \forall k \in \{1,\dots,B\}$, then \Cref{eqn12} can be rewritten as
\begin{equation}\label{eqn13}
  \nabla \boldsymbol{\mathcal{W}}_i = \left( \frac{1}{B} \sum_{k=1}^{B} \frac{\partial{\mathcal{L}(k)}}{\partial{y_i(k)}} \right) \cdot \boldsymbol{e}^\top
\end{equation}
Combining \Cref{eqn8} and \Cref{eqn13}, we have
\begin{equation}\label{eqn14}
    \nabla \boldsymbol{\mathcal{W}}_i = \nabla{b}_i \cdot \boldsymbol{e}^\top
\end{equation}

For the convenience of subsequent discussions, $\nabla \boldsymbol{\mathcal{W}}^{sa}$, $\nabla \boldsymbol{b}^{sa}$ are used to denote the aggregated gradients of weight and bias, $\nabla \boldsymbol{\mathcal{W}}^{u}, \nabla \boldsymbol{b}^{u}, \boldsymbol{e}^{u}$ are used to denote the gradients of weight, the gradients of bias, and the embeddings that come from the client $u$.
And $\nabla \boldsymbol{\mathcal{W}}^{u}_i, \nabla {b}^{u}_i$ are used to denote the $i^{th}$ row of $\nabla \boldsymbol{\mathcal{W}}^{u}$, the $i^{th}$ element of $\nabla \boldsymbol{b}^{u}$.
Here, we suppose that the server has obtained the $\boldsymbol{e}$ of each client and aims to disaggregate $\nabla{\boldsymbol{b}}^{sa}$ into 
$\{\nabla{\boldsymbol{b}}^1,\nabla{\boldsymbol{b}}^2,\dots,\nabla{\boldsymbol{b}}^U \}$, where $\nabla{\boldsymbol{b}}^{sa}$ denotes the aggregated gradients of FCL's bias.

From \Cref{sec:sa}, we have known that the aggregated gradients encrypted by SA are the sum of individual gradients from all selected clients. That is, 
\begin{equation}\label{eqn15}
    \nabla \boldsymbol{\mathcal{W}}^{sa}_i = \sum_{u=1}^{U}\nabla \boldsymbol{\mathcal{W}}^u_i, \; \nabla{b}^{sa}_i = \sum_{u=1}^{U}\nabla{b}^u_i
\end{equation}
Specifically, based on \Cref{eqn14}, we have
\begin{equation}\label{eqn16}
    \nabla \boldsymbol{\mathcal{W}}^{sa}_i = \sum_{u=1}^{U}\nabla{\boldsymbol{\mathcal{W}}^u_i} =\sum_{u=1}^{U} (\nabla{b}^u_i \cdot {\boldsymbol{e}^u}^\top).
\end{equation}
We note that $\nabla{b}^u_i$ is a scalar, while $\boldsymbol{\mathcal{W}}_i^{sa} $ and ${\boldsymbol{e}^u}^\top$ are $m$-dimension vectors. Specifically, ${\boldsymbol{e}^u}^\top = [e^u_1, \dots, e^u_m]$, and $\boldsymbol{\mathcal{W}}_i^{sa} = [w^{sa}_{i,1}, \dots, w^{sa}_{i,m}]$. Hence, we can derive the gradients $[\nabla{b}^1_i,\ldots,\nabla{b}^U_i]^{\top}$ (i.e., gradients of bias related to label $i$) by solving a system of linear equations (linear system) based on \Cref{eqn15} and \Cref{eqn16}, which can be shown as

\begin{equation}\label{eqn17}
\left\{
\begin{aligned}    
    \nabla{b}^1_i + \nabla{b}^2_i + \dots + \nabla{b}^U_i &= \nabla{b}^{sa}_i\\
    \nabla{b}^1_i \cdot {e^1_1} + \nabla{b}^2_i \cdot {e^2_1} + \dots + \nabla{b}^U_i \cdot {e^U_1} &= \nabla{w}^{sa}_{i,1}\\
    \nabla{b}^1_i \cdot {e^1_2} + \nabla{b}^2_i \cdot {e^2_2} + \dots + \nabla{b}^U_i \cdot {e^U_2}& = \nabla{w}^{sa}_{i,2}\\
    \dots\\
    \nabla{b}^1_i \cdot {e^1_m} + \nabla{b}^2_i \cdot {e^2_m} + \dots + \nabla{b}^U_i \cdot {e^U_m} &= \nabla{w}^{sa}_{i,m}\\
\end{aligned}
\right.
\end{equation}

The above linear system can be solved by Moore-Penrose pseudo-inverse algorithm. Moreover, we let $U \leq (m+1)$ to ensure that \Cref{eqn17} has a unique solution, where $U$ is the number of selected clients participating in FL. In this way, we can accurately derive the individual gradients concerning the bias of FCL. Note that the gradients of bias related to all labels can be obtained by repeatedly solving \Cref{eqn17} $n$ times by solely modifying the coefficient matrix. So far, we have obtained the gradients w.r.t FCL's bias of all clients.

\emph{Step 2: Label Inference.}
In \emph{step 1}, we successfully obtain the individual gradients of FCL's bias $\{\nabla{\boldsymbol{b}}^1,\dots,\nabla{\boldsymbol{b}}^U \}$. 
Recall Eq. (\ref{eqn9}), suppose that we obtain the logits $\{\boldsymbol{y}^1,\dots, \boldsymbol{y}^U\}$ from all selected clients. Specifically, $\boldsymbol{y}^u(k)$ $=$ $[y^u_1(k), \dots, y^u_n(k)]$ and $\boldsymbol{y}^u(k) \in \boldsymbol{y}^u$ is the output of FCL (i.e., logits) regarding sample $k$ from client $u$.
We can infer the labels from each client as follows:
\begin{equation}\label{eqn18}
    \lambda^u_i = \sum_{k=1}^{B}\frac{e^{y^u_i(k)}}{\sum_{j=1}^{n}e^{y^u_j(k)}} - B\nabla{b^u_i},
\end{equation}
where $B=|\mathcal{B}^u|$ is specified by the server and $\lambda^u_i$ is exactly the number of label $i$ in the batch $\mathcal{B}^u$ of client $u$. 
By computing \Cref{eqn18} $n \times U$ times, we can recover the label information from all clients. We illustrate the label inference from aggregated gradients in Proposition 3.

\begin{prop}
  The number of occurrences $\lambda_i^u$ of label $i$ in the training batch $\mathcal{B}^u$ of client $u$ can be derived from the aggregated gradients $\nabla{\boldsymbol{\mathcal{W}}}^{sa}$ and $\nabla{\boldsymbol{b}}^{sa}$ when obtaining FCL's input vector (embeddings) $\boldsymbol{e}$  and output vector (logits) $\boldsymbol{y}$ of all clients. 
\end{prop}

To infer labels from the aggregated gradients, Proposition 3 relies on three key assumptions: \emph{Assumption 1}, the server should obtain the embeddings and logits of all clients; \emph{Assumption 2}, the samples in the batch $\mathcal{B}^u$ of client $u$ should produce the same embeddings, i.e., $\boldsymbol{e}^u(k) = \boldsymbol{e}^u$ for $\forall k \in \{1,\dots,B\}$; \emph{Assumption 3}, the embeddings of different clients should different from each other, i.e., $\boldsymbol{e}^u \neq \boldsymbol{e}^v$ if $u \neq v$ for $\forall$  $u,v \in \mathcal{U}$. However, neither of these assumptions can be fulfilled in common FL scenarios. To this end, we assume a malicious server who can modify the global model and attempt to recover labels from the aggregated gradients. Next, we introduce how to design fishing models to meet these assumptions so as to achieve the purpose of label inference.

\subsection{Design of Fishing Model}
In this subsection, we introduce how to modify the global model to design fishing models. The basic idea is to make the embeddings and logits data-agnostic and only determined by the model parameters through modifying the parameters of one batch normalization (BN) layer in the original global model. It should be emphasized that in addition to the BN layers, other fully connected layers or convolutional layers in the model can also be used to design fishing models.

Let $\boldsymbol{W}_i$ and $\boldsymbol{b}_i$ denote the weight and bias of layer $i$, the forward pass on layer $i$ can be shown as $\mathcal{F}_i(z_i) = \boldsymbol{W_i}\otimes z_i + \boldsymbol{b}_i$, where $z_i$ represents the input of layer $i$. Hence, the forward pass $\mathcal{F}$ of a deep neural network composed of multiple layers can be expressed as
\begin{equation}\label{eqn19}
    \mathcal{F}(z_1) = \mathcal{F}_{n}(\dots(\mathcal{F}_2(\mathcal{F}_1(z_1)))).
\end{equation}
Moreover, the type of layer $i$ can be the convolutional layer, the fully connected layer, the BN layer, etc. The operator $\otimes$ is used to abstract the calculation of the  $\boldsymbol{W}_i$ on the input $z_i$, which changes with the type of the layer. Since the output of layer $i$ is also the input of layer $i+1$, we can fix the outputs of the latter layers if we can fix the output of a specific layer. This can be achieved by setting the weight of any layer with a bias to zero, and then the output of that layer is completely determined by the parameters of the bias. To ensure stealthiness, we select the BN layer as the target layer to be modified, where the number of parameters is relatively less than that of other layers.

\textbf{BN Modification}.
The BN layer is widely used in common models to address the issue of shifting data distribution in the middle layers during model training.
Assuming that the $k^{th}$ layer of the model is a BN layer, then the operation that implements batch normalization is
\begin{equation}\label{eqn20}
   \mathcal{F}_k(\hat{z}_k) = \gamma \cdot \hat{z}_k + \beta,
\end{equation}
where $\hat{z}_k$ is the normalized value of $z_k$,  and $\gamma$, $\beta$ is the parameters of the BN layer. The dimension $d$ of parameters  $\gamma$, $\beta$ depends on the number of input features of layer $i$. 
We can fix the output of the BN layer by setting the parameters $\gamma = \mathbf{0}_d$ and $\beta = \mathbf{C}_d$, where every element has a constant value $C \in \mathbb{R}$  and $C>0$. That is, the output of layer $k$ (i.e., BN layer) is always $\mathbf{C}_d$, which is the input of layer $k+1$, and thus produce fixed embeddings and logits (i.e., $\boldsymbol{e} = z_{n}$ and $\boldsymbol{y} = \mathcal{F}_n(z_{n})$). In this context, the \emph{Assumption 1} and \emph{Assumption 2} have been satisfied. To achieve the \emph{Assumption 3}, we just let the server set different $C$ for different clients to ensure that different clients produce different $\boldsymbol{e}$ and $\boldsymbol{y}$. After the BN modification, the original global model transforms into the fishing model.

\textbf{Attack Process}. Based on the method of designing the fishing model, we summarize the attack process LIA-SA into the following three steps.

(1) \emph{Construction of Fishing Models}. The server first constructs $U$ fishing models for $U$ clients.
Specifically, fixing the parameters of other layers in the original model, the server designs each fishing model $F^u$ for client $u$ by setting elements in the BN layer's $\gamma$ to $0$ and that in the $\beta$ to $C^u$, where  
$C^u \neq C^v$ for $\forall u, v \in \mathcal{U}$ and $u \neq v$. Moreover, the server computes the corresponding embeddings $\boldsymbol{e}^u$ and logits $\boldsymbol{y}^u$ by performing a forward pass of each fishing model $F^u$ from random data. The server then sends these fishing models to the clients, who honestly perform one round of training and send back the encrypted gradients to the server.

(2) \emph{Gradient Disaggregation}.  After receiving the encrypted gradients from all selected clients, the server decrypts the gradients using the SA decoder to get the aggregated gradient $\nabla \boldsymbol{\mathcal{W}}^{sa}$, $\nabla \boldsymbol{b}^{sa}$. According to \Cref{eqn17}, the server can build a system of equations based on the embeddings $\{\boldsymbol{e}^1,\dots, \boldsymbol{e}^U\}$ and aggregated gradients  $\nabla \boldsymbol{\mathcal{W}}^{sa}$, $\nabla \boldsymbol{b}^{sa}$ to get $\{\nabla\boldsymbol{{b}}^1, \dots, \nabla\boldsymbol{{b}}^U\}$.

(3) \emph{Label Inference}. Finally, the server employs preset logits $\{\boldsymbol{y}^1,\dots,\boldsymbol{y}^U\}$ obtained by fishing models and individual gradients $\{\nabla\boldsymbol{{b}}^1, \dots, \nabla\boldsymbol{{b}}^U\}$ obtained by gradient disaggregation to infer the number of occurrences of each label in the batch for each client (i.e., \Cref{eqn18}). That is, the server derives the $\{\boldsymbol{\lambda}^1,\dots, \boldsymbol{\lambda}^U\}$, where $\boldsymbol{\lambda}^u = [\lambda_1^u,\dots, \lambda_n^u] (1 \leq u \leq U)$ denotes the label information of client $u$.

\section{Experiments}\label{sec:exp}
In this section, we conduct extensive experiments on various datasets and models to evaluate the performance of LIA-SA in restoring labels.
We first provide a comparison with prior label inference attacks.
Then we analyze the stealthiness of GIAs against SA and show the runtime of LIA-SA.
Finally, we present the performance of LIA-SA against two defense strategies (i.e., differential privacy and gradient compression).
\subsection{Setups}
\textbf{Datasets and Models}\quad We use the following image datasets with ascending classes in our experiments.
\begin{itemize}
    \item CelebA\cite{liu2015faceattributes} contains 202K facial images from more than 10k celebrities. We resize these images to 128$\times$128 and consider only the gender attribute with 2 classes.
    \item MNIST\cite{lecun1998gradient} contains 70k gray-scale handwritten digit images of size 28$\times$28 with 10 classes.
    \item CIFAR10/100\cite{krizhevsky2009learning} contains 60k color images of size 32$\times$32 with 10/100 classes.
    \item ImageNet ILSVRC 2012\cite{deng2009imagenet} contains more than 14M color images of size 224$\times$224 with 1000 classes.
\end{itemize}

 We use a wide range of models including FCN-3 consisting of three fully connected layers, VGG series, and ResNet series.
Since there is no BN layer in FCN-3, we construct the fishing models by modifying the parameters of the first fully connected layer.

\textbf{Implementation Details}\quad
We consider an FL system with $N = 100$ clients in total and $U=5$ clients are randomly selected to participate in each training round by default. For each client, it randomly selects $B$ samples from the training dataset, where the batch size $B=64$ by default. The clients train the models using SGD optimizer with learning rate $0.1$ and cross-entropy loss.
To remove randomness, all the presented statistics are the averaged values of 20 independent experiments.
Furthermore, we conduct our experiments using PyTorch 1.11.0 on a workstation equipped with Intel(R) Xeon(R) Gold 6248R CPU @ 3.00GHz, 95GB RAM, and an NVIDIA A100 (80G) GPU.

\textbf{Evaluation Metrics}\quad
To estimate the performance of LIA-SA, we use the metric \emph{Label Number Accuracy} (LnAcc) \cite{ma2022instance} to represent the accuracy for predicting the number of samples per class (i.e., the number of label occurrences).
Specifically, we redesign LnAcc into the following two metrics:
\begin{itemize}
    \item LnAcc-all: the LnAcc of labels aggregated by all clients.
    \item LnAcc-target: the LnAcc of labels from a target client (can be any client).
\end{itemize}
In \Cref{sec:stealth}, we use the following metrics to evaluate the stealthiness of LIA-SA and other GIAs against SA. 
\begin{itemize}
    \item NoMP: the number of modified parameters.
    \item Ratio: the ratio of the modified parameters to the total model parameters.
    \item CosSim: the cosine similarity between the gradients generated during the attack and the gradients generated during benign training.
\end{itemize}

\begin{figure*}[!t]
    \centering
    \subfigure[VGG-11 on CIFAR10]{
    \includegraphics[width=0.48\columnwidth]{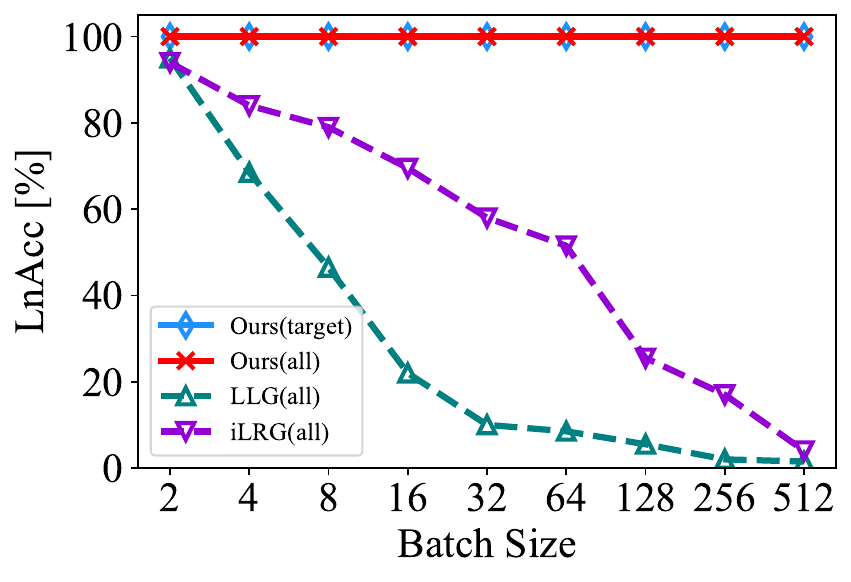}}
    \subfigure[ResNet-18 on CIFAR100]{
    \includegraphics[width=0.48\columnwidth]{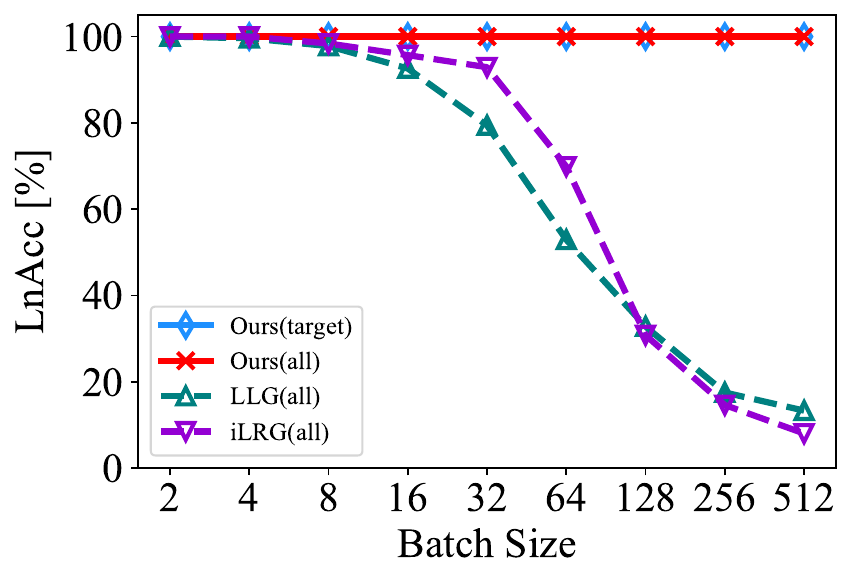}}
    \subfigure[VGG-11 on CIFAR10]{
    \includegraphics[width=0.48\columnwidth]{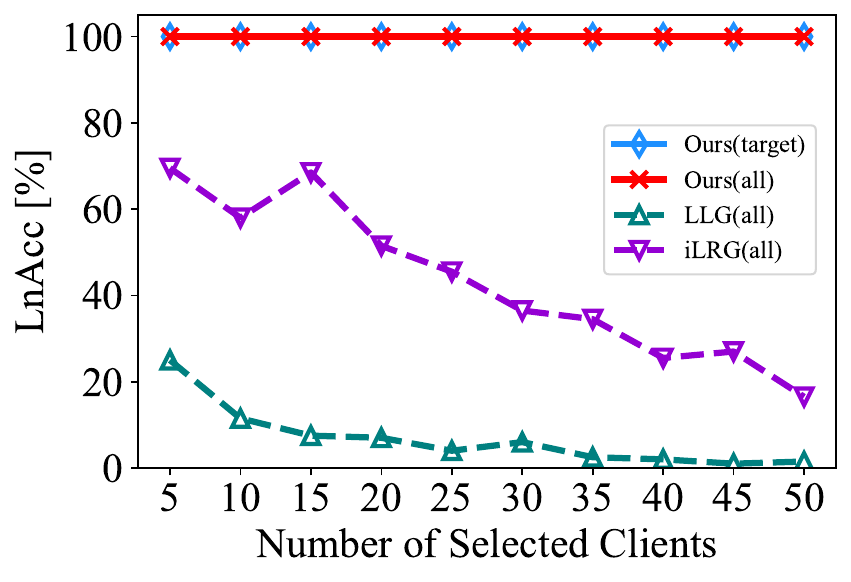}}
    \subfigure[ResNet-18 on CIFAR100]{
    \includegraphics[width=0.48\columnwidth]{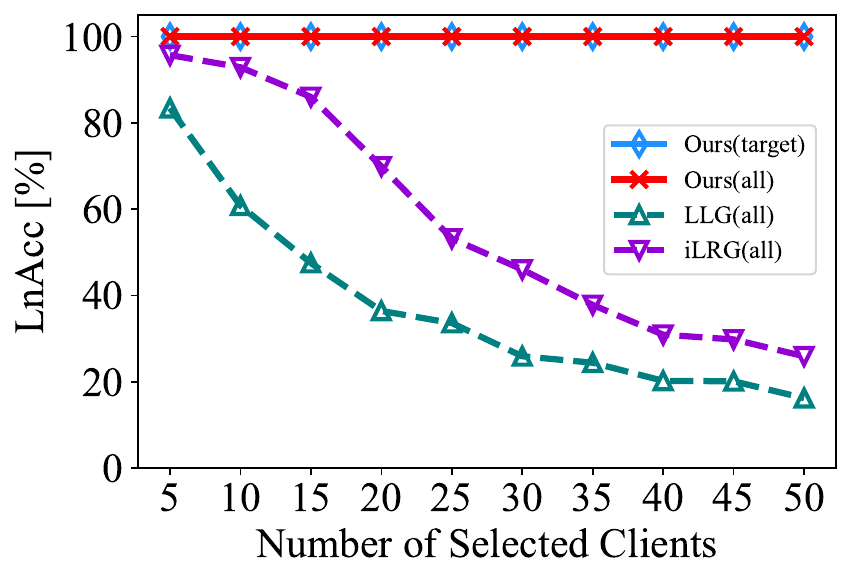}}
    
    \caption{The effect of batch size and number of selected clients on the LnAcc of LIA-SA(ours), LLG, and iLRG. }
    \label{fig:effect_of_bs_con}
    \vspace{-4mm}
\end{figure*}

\textbf{Baselines}\quad We compare our work with the following baselines. The first two baselines (label inference attack) are used to compare the performance on label recovery accuracy, and the last two baselines (GIAs against SA) are used to compare the performance on stealthiness.    
\begin{itemize}
\item LLG\cite{wainakh2021user} exploits the direction and magnitude of gradients to infer labels from the individual gradients.
\item iLRG\cite{ma2022instance} recovers the instance-wise batch label from the individual gradients based on the approximate recovered class-wise embeddings and post-softmax probabilities.
\item EludingSA\cite{pasquini2022eluding} reconstructs the input data against SA by modifying the model parameters to make the non-target clients generate zero gradients, which helps to obtain the individual gradients of the target user.
\item MANDRAKE\cite{zhao2023secure} reconstructs the input data against SA by adding a convolutional layer and two fully connected layers in front of the original model to separate gradients from different clients and different samples.
\end{itemize}

\subsection{Comparison with Prior Label Inference Attacks}

\begin{table}[!htbp]
    \small
    \caption{Comparison of LIA-SA with prior works on different datasets.}
    \centering
    \tabcolsep=1pt
    \renewcommand{\arraystretch}{1.1}
    \begin{threeparttable}
    \begin{tabular}{cccccccc} 
    \toprule 
    \multirow{2}{*}{Datasets(class)}& \multirow{2}{*}{Model}& \multicolumn{2}{c}{LLG}& \multicolumn{2}{c}{iLRG}& \multicolumn{2}{c}{LIA-SA(ours)}\\
    \cmidrule{3-8}
    ~&~&all*&target&all&target&all&target\\  
    \hline 
    MNIST(10)& FCN-3&0\%&-&2.00\%&-&100\%&100\%\\
    \hline
    CelebA(2)& ResNet-18&0\%&-&5.00\%&-&100\%&100\%\\
    \hline
    CIFAR10(10)& VGG-11&3.00\%&-&24.50\%&-&100\%&100\%\\
    \hline
    CIFAR100(100)& ResNet-18&1.50\%&-&30.30\%&-&100\%&100\%\\
    \hline
    ImageNet(1000)& ResNet-50&9.31\%&-&98.54\%&-&100\%&100\%\\
    \bottomrule 
    \end{tabular}
    \begin{tablenotes}
        \footnotesize
        \item * all and target are abbreviations for LnAcc-all and LnAcc-target.
    \end{tablenotes}
    \end{threeparttable}
    \label{table:compare_with_prior}
\end{table}

In this subsection, we evaluate the performance of LIA-SA and two existing label inference attacks \cite{wainakh2021user,ma2022instance} on label restoration under SA setting. 
Note that LLG and iLRG can only work on the aggregated gradients and are unable to recover the labels of a target client.  Therefore we only use the metric LnAcc-all  to evaluate these two attacks.

\begin{figure}[!t]
    \vspace{-2mm}

    \centering
    \subfigure[VGG-series on CIFAR100]{\includegraphics[width=0.49\linewidth]{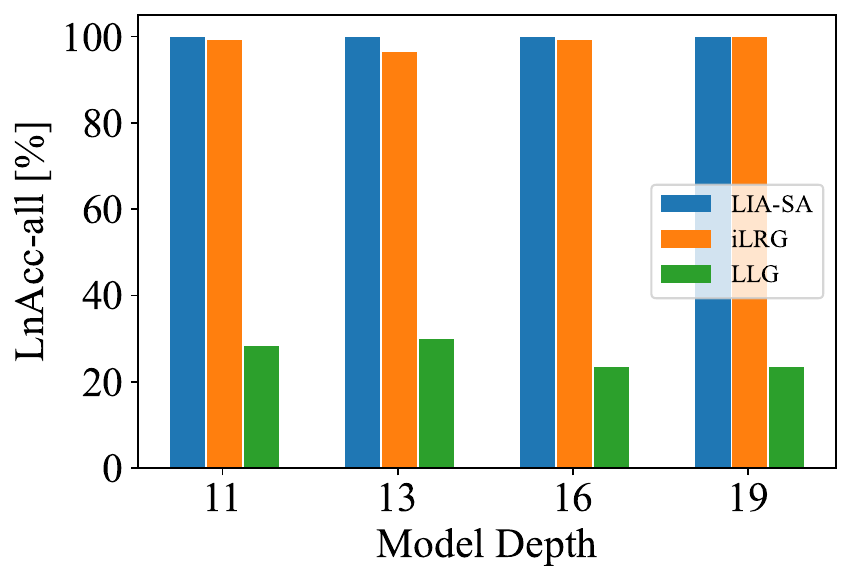}}
    \subfigure[ResNet-series on CIFAR100]{\includegraphics[width=0.49\linewidth]{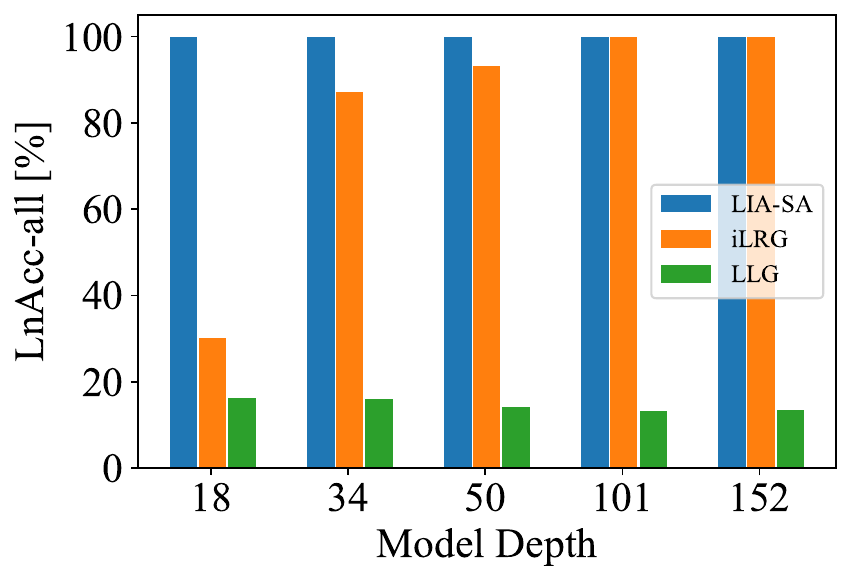}}
    \caption{The effect of model depth on the label inference attacks.}
    \label{fig:model_depth}
    \vspace{-4mm}
\end{figure}

\emph{1) Effect of Batch Size and Number of Selected Clients:} Here, we first show how two FL parameters, batch size and number of selected clients, affect the performance of label inference attacks.
We conduct experiments on CIFAR10 and CIFAR100 with VGG-11 and ResNet-18, respectively.
From \Cref{fig:effect_of_bs_con}, we can observe that the LnAcc-all of LLG and iLRG decreases dramatically as the batch size and the number of selected clients increase.
On the contrary, our method consistently achieves 100\% accuracy across a range of settings.
This demonstrates that LIA-SA would not be affected by the batch size and the number of selected clients.
We should note that the number of selected clients is not unlimited and it should be less than the dimension of embeddings (i.e. $U \leq m+1$) to ensure a unique solution for label inference.

\emph{2) Effect of Dataset and Model:} We then evaluate the performance of our proposed LIA-SA and other label inference attacks on different datasets and model depths. The batch size is 1024 and the number of selected clients is 5 in this section. Tab.~ \ref{table:compare_with_prior} shows the effect of different datasets on the label inference attacks.
LLG performs poorly in all cases, suggesting that it is hard for LLG to infer useful label information from aggregated gradients generated from a large number of samples (e.g., $1024\times5$).
For iLRG, it shows a good performance on datasets with large classes (e.g. ImageNet) but performs poorly in other cases. This is because repeat labels in one batch hugely harm the performance of iLRG.
However, our method always achieves 100\% accuracy for both LnAcc-all and LnAcc-target in all scenarios. Our method not only accurately infers labels from all clients but also individual labels from the target client.

We also evaluate the effect of model architecture and depth on the performance of LIA-SA, iLRG, and LLG using VGG-series and ResNet-series.
As shown in \Cref{fig:model_depth},
the accuracy of LLG is below 30\% under various model depths. iLRG performs well on the VGG-series and the accuracy can reach 99\%.
As for ResNet-18, the performance of iLRG drops significantly with only 30\% accuracy, while it achieves 100\% accuracy on ResNet-152. This shows that the performance of iLRG is affected by both model architecture and model depth.
However, our attack consistently achieves 100\% accuracy on various model architectures and depths, indicating it is efficient and model-agnostic. This implies that our attack can be applied to other models beyond those mentioned in this paper, making it a versatile and powerful method for label recovery in FL.

\begin{table}[!t]
    \centering
    \tabcolsep=2pt
    \renewcommand{\arraystretch}{1.2}
    \caption{Stealthiness Evaluation of GIAs against SA.}
    \begin{tabular}{ccccc} 
    \toprule 
    Attack& Model & NoMP $\downarrow$& Ratio $\downarrow$&CosSim$\uparrow$\\
    \hline
    \multirow{2}{*}{EludingSA\cite{pasquini2022eluding}}&VGG-11&$128$&$4.55\times10^{-4}\%$&0.0002\\
    ~&ResNet-18&$128$&$1.14\times10^{-4}\%$&0.062\\
    \hline 
    \multirow{2}{*}{MANDRAKE\cite{zhao2023secure}}&VGG-11&$8.65\times10^6$&$23.51\%$&0.624\\
    ~&ResNet-18&$8.65\times10^6$&$43.53\%$&0.752\\
    \hline 
    \multirow{2}{*}{Ours}&VGG-11&$128$&$4.55\times10^{-4}\%$&0.981\\
    ~&ResNet-18&$128$&$1.14\times10^{-4}\%$&0.968\\
    \bottomrule 
    \end{tabular}
    \label{tabel:stealthiness}
    
    \vspace{-4mm}
\end{table}

\subsection{Stealthiness Evaluation of GIAs against SA}\label{sec:stealth}

Stealthiness reflects the difficulty of the attack being detected and is hard to measure. In this subsection, we evaluate the stealthiness of LIA-SA and some existing GIAs against SA \cite{pasquini2022eluding,zhao2023secure}  on proposed metrics NoMP, Ratio, and Cossim. 
The used dataset is CIFAR100,  and the models are VGG-11 and ResNet-18.
As shown in Tab. \ref{tabel:stealthiness},
EludingSA and LIA-SA modify only a small number of parameters of the BN layer and perform well on the NoMP and Ratio.
However, EludingSA induces non-target clients to generate zero-gradient, which results in low CosSim, while the gradients generated by LIA-SA are similar to the original gradients, thus exhibiting high CosSim.
Besides, MANDRAKE makes modifications to the victim's model and adds millions of parameters to the original model, thus performing badly on NoMP, Ratio, and Cossim.  On the contrary, LIA-SA does not modify the model architecture and only modifies a few parameters in BN layers.
In summary, our method modifies trivial parameters and has high gradient similarity, ensuring better stealthiness than EludingSA and MANDRAKE.

\subsection{Runtime Evaluation}
\begin{figure}[!t]
\centering
\subfigure[]{\includegraphics[width=0.49\linewidth]{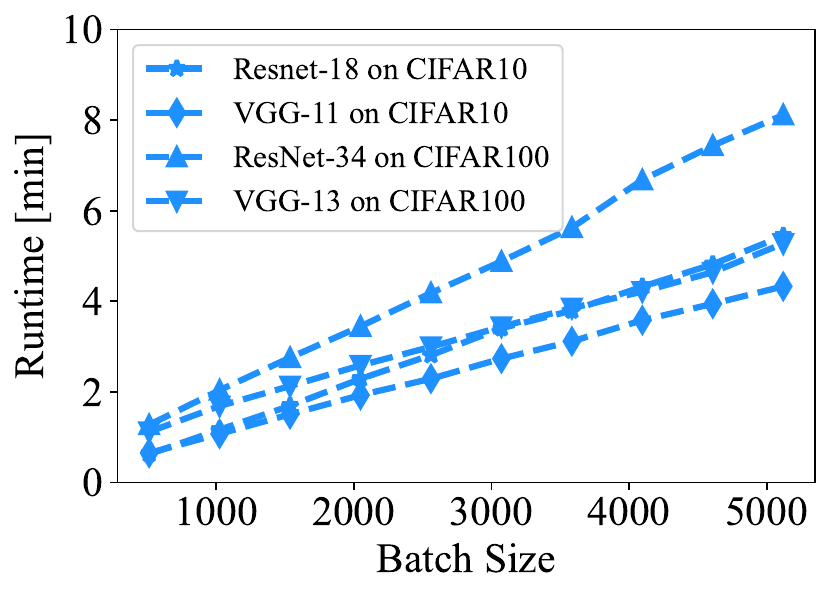}} 
\subfigure[]{\includegraphics[width=0.49\linewidth]{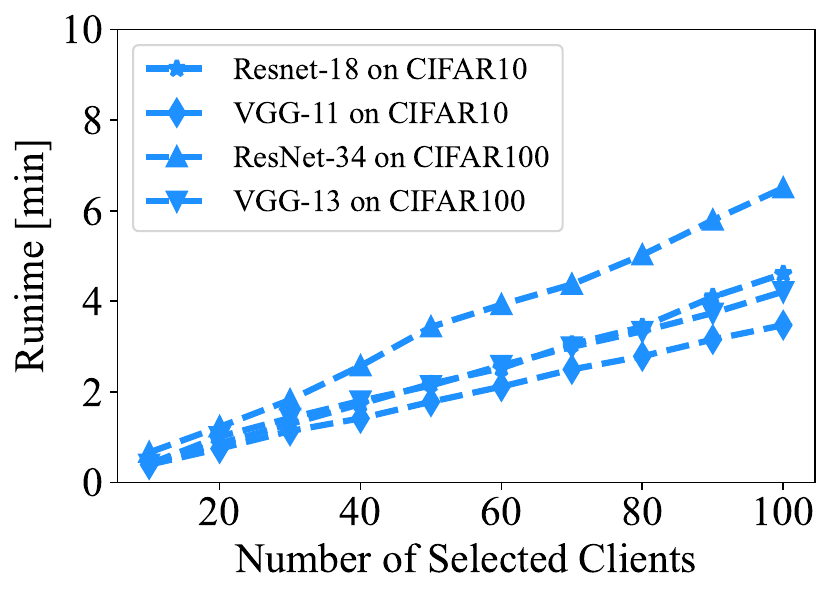}}
\caption{Runtime evaluation of LIA-SA with varying batch size and the number of selected clients.}
\label{fig:efficiency_evaluation}
\vspace{-1mm}
\end{figure}

This subsection shows the runtime of LIA-SA to illustrate the computation cost of LIA-SA.
In this experiment, we consider larger batch size and number of selected clients. We increase the number of selected clients to 100 and the batch size to 5120 constrained by our device. The runtime of LIA-SA regarding the batch size and number of selected clients is shown in \Cref{fig:efficiency_evaluation}.
It can be observed that when the batch size and the number of selected clients are small, LIA-SA can be completed within 1 minute.
Even in a relatively large-scale FL setting with a batch size of 5120 and a number of selected clients of 100, the runtime is only 8 minutes.
Moreover, the most important observation is that the runtime linearly changes related to the batch size and the number of selected clients,
which indicates that LIA-SA can achieve large-scale label inference (i.e., a significantly large number of clients and batch size) with enough computing resources.

\subsection{Performance of LIA-SA against Defense Strategies}
\begin{figure}[!t]
    \vspace{1mm}
    \centering
    \subfigure[Differential Privacy]{\includegraphics[width=0.49\linewidth]{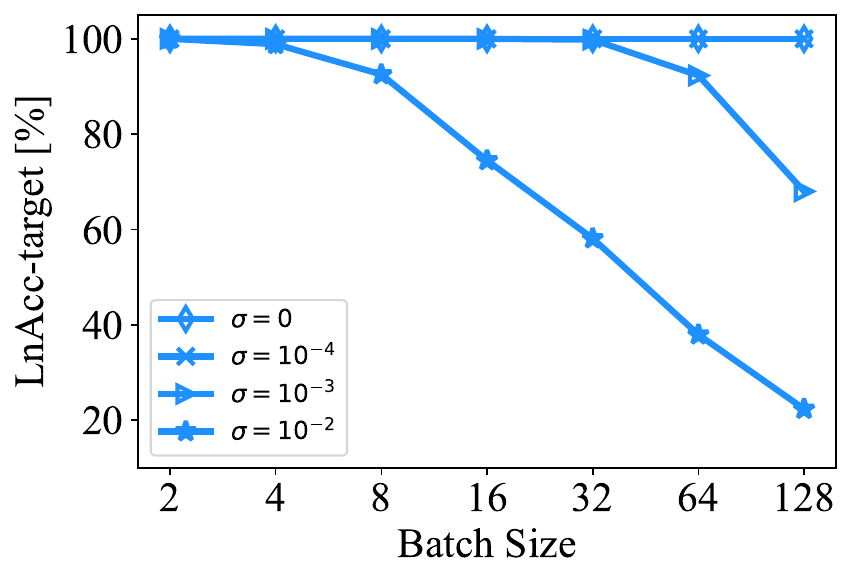}}
    \subfigure[Gradient Compression]{\includegraphics[width=0.49\linewidth]{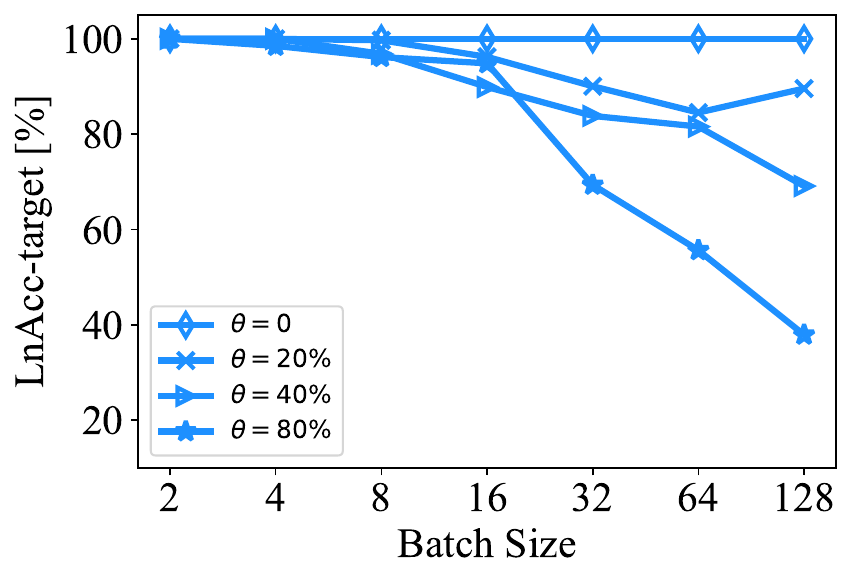}}
    \caption{The effect of Differential Privacy and Gradient Compression on LIA-SA.
    $\sigma = 0$ or $\theta = 0$ represents no additional defense.}
    \label{fig:defense}
    \vspace{-1mm}
\end{figure}
Our attack relies on the gradient information both in gradient disaggregation and label inference, so some gradient obfuscation-based defense strategies may degrade the performance of our attack. These defense strategies are deployed on the client side, who will process the gradients before sending them to the server.
In this subsection, we explore the impact of the following two gradient obfuscation-based defense mechanisms on LIA-SA: 
(1) Differential Privacy (Additive Noise) \cite{dwork2006differential}: adding a  Gaussian noise $\epsilon \sim \mathcal{N} (0,\sigma^2\boldsymbol{I})$ to the gradients.
In our evaluation, we set the standard deviation of the noise distribution as $\sigma = \{0,10^{-4}, 10^{-3}, 10^{-2}\}$.
(2) Gradient Compression \cite{lin2017deep}: pruning the gradients less than a certain threshold to 0 and keeping the gradients that are larger than a certain threshold.
We evaluate the defense effect with different compression ratios $\theta=\{0,20\%,40\%,80\%\}$. The evaluation is performed on CIFAR10 with ResNet18.
From \Cref{fig:defense}, we can observe that when the standard deviation of the noise distribution is not particularly large (i.e., $\sigma \leq 10^{-3}$),
LIA-SA still maintains high accuracy. 
For gradient compression, our attack performs better with a small batch size and still achieves 90\% accuracy even with an 80\% compression ratio when the batch size is 16.
This demonstrates that LIA-SA exhibits a certain level of resilience against the defense of gradient obfuscation.

\section{Conclusion and Future Work}\label{con}
In this work, we propose LIA-SA, a label inference attack to bypass SA and steal label information from aggregated gradients in FL.
Inspired by the analysis of label inference from aggregated gradients, we design the fishing model to obtain clients' embeddings and logits.
Based on the embeddings of all clients and the aggregated gradients,
we build linear equations to disaggregate the aggregated gradients of FCL into individual gradients.
Then the labels of each client can be computed based on the individual gradients of FCL's bias and logits.
Compared to prior works, we only modify trivial model parameters in a single BN layer and achieve good stealthiness.
Extensive experiments show that LIA-SA can recover the labels of each client from the aggregated gradients with 100\% accuracy on various datasets and model architectures. However, LIA-SA only focuses on recovering label information from aggregated gradients under FedSGD.
 In the future, we aim to focus on FedAvg and investigate the possibility of stealthily recovering data information from the aggregated gradients. This direction will expand our understanding of the privacy vulnerabilities in FL and SA, exploring potential defense mechanisms against GIAs.

\section*{Acknowledgments}
This work was supported by National Key R\&D Program of China (Grant No. 2021ZD0112803), National Natural Science Foundation of China (Grants No. 62122066, U20A20182, 61872274) and  Key R\&D Program of Zhejiang (Grant No. 2022C01018).


\bibliographystyle{IEEEtran}
\bibliography{reference}

\end{document}